\documentclass{aa}

\usepackage{psfig,latexsym,amssymb,times}

\begin{document}
\title{The BL Lac objects OQ 530 and S5~0716+714}
\subtitle{Simultaneous observations in the X--rays, radio, optical and TeV bands}
\author{G.~Tagliaferri\inst{1}
\and M. Ravasio\inst{1}
\and G. Ghisellini\inst{1}
\and P. Giommi\inst{2}
\and E. Massaro\inst{3}
\and R. Nesci\inst{3}
\and G. Tosti\inst{4}
\and M.F. Aller\inst{5}
\and H.D. Aller\inst{5} 
\and A. Celotti\inst{6}
\and L. Maraschi\inst{7}
\and F. Tavecchio\inst{1}
\and A. Wolter\inst{7}
}
\offprints{G. Tagliaferri (gtagliaf@merate.mi.astro.it)}

\institute{INAF -- Osservatorio Astronomico di Brera, Via Bianchi 46, I-23807 Merate, Italy 
\and ASI Science Data Center, Via Galileo Galilei, I-00044 Frascati, Italy
\and Dipartimento di Fisica, Universit\`a La Sapienza, P.le Aldo Moro 2, I-00185 Roma, Italy
\and Dipartimento di Fisica e Osservatorio Astronomico, Universit\`a di Perugia, Perugia, Italy
\and Department of Astronomy, University of Michigan, Dennison Building, Ann Arbor MI
     48109-1090, USA
\and SISSA/ISAS, via Beirut 2-4, I-34014 Trieste, Italy
\and INAF -- Osservatorio Astronomico di Brera, Via Brera 28, I-20121 Milano, Italy 
}

\date{Received October 2002}



\abstract{
We present the results of the $Beppo$SAX observations of two 
BL Lacs, OQ\,530 and S5\,0716+714, as part of a ToO program for
the simultaneous observation at radio, optical, X--ray and TeV energies.  
Both sources are detected in the LECS and MECS, with 
S5\,0716+714  visible also in the PDS band, up to about $\sim$ 60 keV. The
X-ray spectra of both sources are better fitted by a double power--law
model, with a steep soft X--ray component flattening at harder
energies, with breaks at 0.3 and 1.5 keV, respectively.  The
concave shape of the spectra in both objects is consistent with soft
X--rays being produced by the synchrotron and harder X--rays by the inverse
Compton processes. Also the X--ray variability properties confirm this
scenario, in particular for S5\,0716+714 our observation shows variations 
by about a factor 3 over one hour
below $\sim$ 3 keV and no variability above.  Their simultaneous
broad band energy spectral distributions can be successfully
interpreted within the frame of a homogeneous synchrotron and inverse
Compton model, including a possible
contribution from an external source of seed photons
with the different spectral states of S5\,0716+714 being
reproduced by changing the injected power.
The resulting parameters are fully consistent with the two sources being 
intermediate objects within the ``sequence" scenario proposed for blazars.
\keywords{
BL Lacetae objects: general -- X-rays: galaxies -- BL Lacetae objects: 
individual: OQ\,530, S5\,0716+714}}

\maketitle

\section{Introduction}
Blazars are radio-loud Active Galactic Nuclei whose spectra 
are characterized by  variable nonthermal continua emitted by 
relativistic jets oriented close to the line of sight.
They display  double peaked Spectral Energy Distributions
(SED) extending from radio up to $\gamma$-rays and sometimes to TeV frequencies.
The first peak (radio to UV/X--rays) is generated by synchrotron emission
produced by relativistic electrons flowing along the jet, while the 
second component (extending to $\gamma$--rays) is commonly attributed 
to inverse  Compton emission. The same synchrotron emitting electrons 
upscatter ``seed'' photons to higher energies. The origin of this seed 
radiation field is still debated: it could be the synchrotron radiation itself
(SSC models, Maraschi, Ghisellini \& Celotti, 1992) or produced outside 
the jets by different mechanisms (ERC, external radiation Compton, see
Sikora, Begelman \& Rees, 1994; Ghisellini \& Madau,1996; Dermer \& 
Schlickeiser, 1993; Blazejowski et al., 2000). Blazars are divided in 
two subclasses according to their spectral features: with respect to 
the presence or the absence of broad emission lines (EW$> 5$ \AA)
they are classified as Flat Spectrum Radio Quasars or as BL Lac objects
respectively.
BL Lacs are further classified as HBL or LBL according to the 
position of the synchrotron peak: LBLs peak 
in optical--UV band  while HBLs in X--rays (Padovani \& Giommi, 1995).
The extremely wide energy range of the SED and the high variability 
of these sources  makes simultaneous multiwavelength observations 
a powerful tool to understand their physics.
Therefore we proposed a number of  Target of Opportunity 
{\it Beppo}SAX observations of blazars in high state of activity, 
coordinated with optical and TeV monitoring campaigns: the brightening of
the source in one of these bands would trigger {\it Beppo}SAX NFI action. 
As part of this program we already presented {\it Beppo}SAX 
observations of ON 231, PKS 2005--489 
(Tagliaferri et al. 2000, 2001) and BL Lac (Ravasio et al. 2002). 
In this paper we report the results of the {\it Beppo}SAX observations
of OQ\,530 and S5\,0716+714 (the fourth and fifth triggers of our program).
 
{\bf OQ\,530 ---} It was  optically identified in 1977 (K\"uhr 1977) 
as counterpart of an extragalactic source included in the 
NRAO-Bonn radio survey at 5 GHz. In 1978, it was classified 
as a BL Lac by Miller (1978), because of its stellar aspect 
and the faintness of its optical spectral features.
The redshift z=0.152 has been determined
from absorption features of the host galaxy as well as from 
[OII], $\lambda=3727$ \AA~ and [OIII], $\lambda\lambda=4959,5007$ \AA~~ emission lines
(Stickel et al. 1991; Stickel et al. 1993), confirming previous estimates
by Hagen--Thorn \& Marchenko (1989). 
The host galaxy of OQ\,530 was studied by Abraham et al. (1991), Stickel et al.
(1993), Wurtz et al. (1996) and Scarpa et al. (2000). The data given by these
authors, transformed into R$_C$ (Cousins) give
slightly different values for the host galaxy magnitude: 16.8, 16.2, 16.6 
and 16.1, respectively . The first three used ground
based instruments and found the galaxy to be probably an S0, the last ones
used HST observations and found the galaxy to be an elliptical, like all 
other known host galaxies of BL Lac sources.
In our monitoring, started in 1994, we never observed OQ\,530 fainter
than $R_C=15.8$, so all these values are formally acceptable for the luminosity of
the host.
Given the systematic uncertainties of these estimates we adopt here an
average value of $R_C=16.5$.
The effective radius of the galaxy is found by all authors to be less than 4",
so that it is nearly fully included in our photometric aperture (radius 5" ).
Our BeppoSAX pointing was made when OQ\,530 was at $R_C=14.8$, so the host 
contribution to the optical flux is not negligible: we estimated this
contribution assuming
the spectral shape and k--correction for an elliptical galaxy by Fukugita et 
al. (1995), i.e. $B=$18.59, $V=$17.17, $R_C=$16.50 and $I_C=$15.74.

In the X--ray band OQ\,530 has been detected
by the {\it Einstein} Observatory in May 1980. The data are well fitted by
a power law spectrum with $\alpha=0.48^{+1.0}_{-0.3}$ in the 0.2-3.5 keV
range and a flux $F_{1 keV}=0.21
\mu$Jy (Worral \& Wilkes 1990).
Some years later, in May 1984, a three days EXOSAT observation
highlighted a non variable behaviour of the source on short time scales,
with a flux of $F_{1 keV} \simeq 0.15\pm0.02 \mu$Jy
(Giommi et al. 1990). ROSAT observed this BL Lac object in
July 1990. The 0.1--2.4 keV spectrum is well fitted 
by a single power law with an energy index 
$\alpha=1.04\pm0.05$ and flux F$_{1 keV}=0.32 \mu$Jy 
(Comastri et al. 1995). {\it Beppo}SAX already observed
OQ\,530 in February 1999. The source was detected by the LECS and MECS
($<10$ keV)  but not by the PDS. A single power law model
fit the data ($\alpha=0.55^{+0.27}_{-0.32}$), but the residuals
towards low energies suggest a concave spectrum (Giommi et al. 2002). 

The optical light curve of OQ\,530 from 1905 until 1977 was derived by Miller
(1978) using the Harvard College plate collection. A further study from 1966
to 1980 was performed by Barbieri et al. (1982) using the Asiago Schmidt plates.
A reanalysis of these works, and an updating of the light-curve until 1997 
was made by Nesci et al. (1997). A collection of published data until 1993
has been recently published by Fan \& Lin (2000). The source shows an
average monotonic decreasing trend 
since the beginning of 1900, when it was at $B\sim$12, until 1995 when its
average value was $B\sim$16, with fast and irregular variations of about 1.5--2 
magnitudes. It is still monitored by some of us (Perugia and Roma groups).
On the night of February 14, 2000, we found the source very bright in
the optical with $R_C=$14.3.
Thus, we triggered the {\it Beppo}SAX TOO, but due to constraints on the
satellite pointing, the source was observed only on March 3, 2000, when 
it had a magnitude of $R_C=$14.9. Moreover, due to problems with the satellite, 
the observation lasted less than 27 ks, instead of the 50 ks allocated.
A second pointing was then performed on March 26, for about another 23 ks;
during this second pointing the $R_C$ magnitude was 14.8. In both cases the 
source was weak in the X--rays and detected only up to 10 keV (see section 2).

{\bf S5\,0716+714 ---} It was discovered in 1979 as the 
optical counterpart of an extragalactic radio source during another
Bonn--NRAO survey (K\"uhr et al. 1981). Two years later it was classified as
a BL Lac by Biermann et al. (1981) because of its featureless optical 
spectrum and high linear polarisation.
It is one of the brightest and most variable BL Lac objects, but, despite 
the many attempts to measure its redshift (the last one by Rector \&
Stoke (2001) with the Keck telescope), only a lower limit
$z > 0.3$ has been estimated by Schalinski et al. (1992) and Wagner et al. 
(1996) on the basis of the non detection of a host galaxy in deep images.

S5\,0716+714 was first detected in the X--ray band by the {\it Einstein}
Observatory, but the flux was too low for a detailed spectral analysis 
(Biermann et al. 1981). In 1991 March 8--11, the source was observed several
times with the ROSAT PSPC detector (0.1--2.4 keV) (Cappi et al. 1994; Urry et
al. 1996; Wagner et al. 1996) and was found to be bright and rapidly variable
with a mean flux of $\sim 10^{-11}$ erg cm$^{-2}$ s$^{-1}$. S5\,0716+714 has
been already observed twice by {\it Beppo}SAX, in 1996 November 14 and in 1998 
November 7 (Giommi et al. 1999)
but the source was not very bright and it was detected only up to 10 keV.
The {\it Beppo}SAX SEDs showed the presence of a steep power law component
below $\sim 2.5$ keV with a spectral index $\alpha_1\sim 1.5$ 
becoming harder towards higher frequencies, with an index $\alpha_2\sim 0.85$.
The former component was also found to be variable in correlation
with the optical flux on a time scale of a few hours (Giommi et al. 1999).

S5\,0716+714 has been regularly monitored in the optical since the end of 1994;
a very bright level was recorded in February 1995 (Ghisellini et al. 1997),
when it reached the magnitude of 12.78 in the $R_C$ bandpass, while in the 
following two years it has never been observed brighter than $R_C=13.0$.
A new strong outburst occurred in September 
1997 when S5\,0716+714 was even brighter than in 1995 at $R_C$=12.58
(Massaro et al. 1999). After a relatively quiet period, in October 2000 
we measured an optical flux comparable to that of  September 1997 and a
{\it Beppo}SAX ToO was activated. Unfortunately, for technical reasons 
the pointing was performed a week after the occurrence of the optical 
maximum and the brightness of S5\,0716+714 had already decreased by about
half a magnitude. In any case, it was brighter than in the two previous 
{\it Beppo}SAX observations, when it was around $R_C$=13.8,
and, for the first time, it was detected in the
hard X--rays by the PDS up to about 60 keV.
\vskip 0.3 true cm

The paper is organised as follows: in section 2 we present the {\it Beppo}SAX
observations and data analysis and in section 3 the radio, optical and TeV 
observations; in section 4 we present the SEDs of the two sources in the
framework of an SSC model. The discussion is given in section 5.

\section{{\it Beppo}SAX Observations and Data Reduction}

{\it Beppo}SAX observed OQ\,530 (1418+546) twice during  2000
(March 3--4, March 26--27).
In both observations the source was not detected by the PDS.

S5\,0716+514 was observed for about one day from October 30
to October 31, 2000; and it was detected up to $\sim 60$ keV.

Standard procedures and selection criteria were applied to the data to avoid
the South Atlantic Anomaly, solar, bright Earth and particle contamination using
the SAXDAS v. 2.0.0 package. Exposures and count rates of the observations 
are reported in Table \ref{tab1}.

\begin{table*}[!t!]
\begin{center}
\begin{tabular}{|l|cc|cc|cc|}
\multicolumn{7}{c} {\bf{OQ\,530}}\\

\hline
 & \multicolumn{2}{|c|} {\bf{LECS}}  & \multicolumn{2}{|c|} {\bf{MECS}} & 
          \multicolumn{2}{|c|} {\bf{PDS}}  \\
\hline
Date & exposure (s) &  count rate$^{\rm a}$ & exposure (s) &  count rate$^{\rm b}$ & 
exposure (s) &  count rate$^{\rm c}$ \\
\hline
3-4 March 2000 & 19625  & $(7.4 \pm 0.8) \times10^{-3}$  & 26633 & $(1.3 \pm 0.1) 
\times10^{-2}$ & 10270 & $(1 \pm 6) \times10^{-2} $ \\
\hline
26-27 March 2000 & 17181 & $(8.2 \pm 0.9) \times10^{-3}$  & 22718 & $(1.1 \pm 0.1) 
\times10^{-2}$   & 10361 & $(2 \pm 6) \times10^{-2}$ \\
\hline

\multicolumn{7}{c}{}\\

\multicolumn{7}{c}{\bf{S5\,0716+714}}\\

\hline

30-31 October 2000 & 19192 & $(4.7 \pm 0.2) \times10^{-2}$ & 43459 & $(5.6 \pm 0.1)
\times10^{-2}$     & 22444 & $(0.12 \pm 0.04) $ \\
\hline

\end{tabular}
\end{center}
\caption{Journal of {\it Beppo}SAX observations. $^{\rm a}$ 0.1--10 keV; 
$^{\rm b}$ 1.5--10 keV; $^{\rm c}$ 12--100 keV. For OQ\,530 the PDS count rates 
are lower than the uncertainties and we excluded these data from our analysis.}
\label{tab1}
\end{table*}

\begin{table*}[!t!]
\begin{center}
\begin{tabular}{|c|c|c|c|c|cc|c|c|}
\multicolumn{9}{c}{\bf{OQ\,530}}\\
\hline
Date & $\alpha_1$ &  K$^*_1$ & $\alpha_2$ & K$^*_2$ & \multicolumn{2}{c|} {$F_{1 keV} ~\mu$Jy} 
  & $F^{**}_{2-10 keV}$ & $\chi^2_r/d.o.f.$ \\
 & & ($\times10^{-4}$) & & ($\times10^{-4}$)  & (1) & (2) & ($\times10^{-12}$) & \\
\hline
3--4 March & $0.55 \pm 0.2$ & $2.1^{+0.6}_{-0.5}$ & & & $0.14$ &  & $1.1$ & $ 1.64/23$\\ 
\hline
3--4 March & $6.7^{+0.2}_{-1.8}$ & $\sim10^{-3}$ & $0.40\pm0.2$ & $1.8 \pm0.5$ & $\sim10^{-4}$ 
    & $0.12$ & $1.2$         & $0.96/21$ \\
\hline
26--27 March &  $ 0.75 \pm 0.20$ & $2.5^{+0.5}_{-0.6}$ &  &  & $0.16$ & & $0.9$ & $1.03/21$\\
\hline
26--27 March &  $6.7 \pm ?$ & $\sim10^{-3}$ &  $0.65 \pm 0.2$ & $ 2.25\pm 0.55$ & $\sim10^{-4}$
    & $0.15$ & $1.0$ &  $0.95/19$\\
\hline

\multicolumn{9}{c}{}\\
\multicolumn{9}{c}{\bf{S5\,0716+714}}\\

\hline
30-31 October, 2000$^a$ & $1.25\pm0.08$ & $16.4\pm1.2$ &  &  &  $1.1$ &  & $2.9$ & $1.99/76$\\
\hline
30-31 October, 2000$^b$ & $2.40^{+0.4}_{-0.3}$ & $10.2\pm0.45$ & $0.60^{+0.25}_{-0.35}$ 
     & $6.2^{+3.3}_{-3.0}$ &  $0.7$ & 0.4 & $3.3$ & $0.92/77$ \\
\hline
\end{tabular}
\caption{X--ray spectral parameters and fluxes.
 $^*$: cts keV$^{-1}$ cm$^{-2}$ s$^{-1}$. $^{**}$: erg cm$^{-2}$ s$^{-1}$. 
The values are obtained performing 
the fit with the absorption parameter fixed at the galactic value 
$N_{\rm H} = 1.18 \times 10^{20}$ cm$^{-2}$ for OQ\,530 and 
$N_{\rm H}=3.9\times20^{20}$ cm$^{-2}$ for S5\,0716+714 (Dickey \& Lockman, 1990). 
$^a$ fit performed only on the LECS+MECS data, $^b$ added also the PDS data to the fit.
The reported uncertainties are the 90$\%$ confidence ranges for one parameter of interest, 
i.e. $\Delta \chi^2=2.7$.
}
\end{center}
\label{table2}
\end{table*}

\begin{figure*}
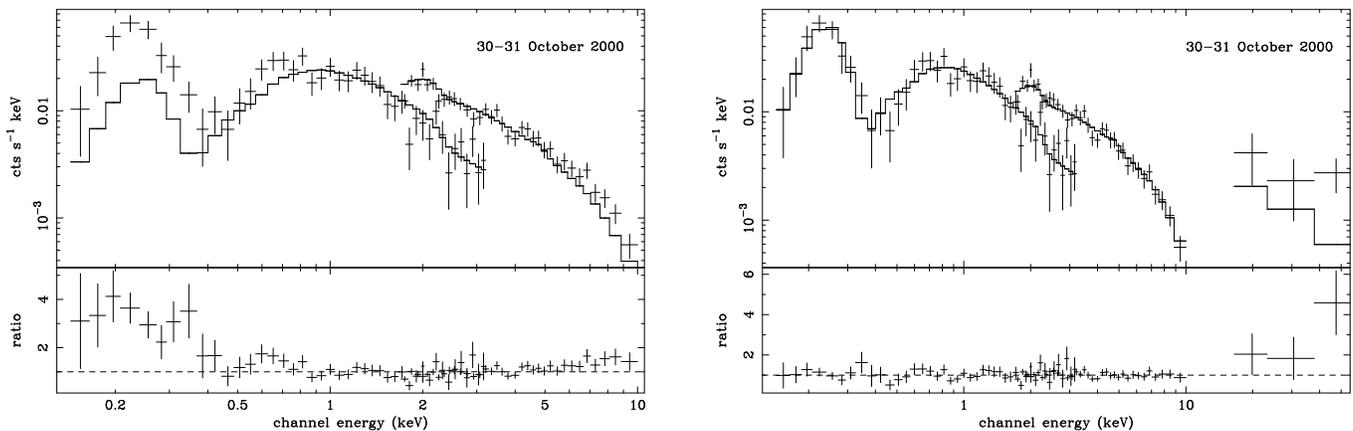

\begin{center}
\hbox to \textwidth{
\centerline{
%
%
\vbox {\psfig{figure=MS3234f1a.ps,angle=-90,width=8.5cm}}
\hskip .8cm
\vbox{\psfig{figure=MS3234f1b.ps,angle=-90,width=8.5cm}}
\hfill }}
\caption{Left panel: LECS+MECS S5\,0716+714 spectrum fitted by a 
single power law model. Positive residuals are evident below $\sim 1$ keV; 
they get larger towards lower energies.
Right panel: LECS+MECS+PDS S5\,0716+714 spectrum fitted by a two 
power law model. In both cases we fitted the data with the absorption 
parameter fixed at the galactic value $N_{\rm H}=3.9\times10^{20}$ cm$^{-2}$ 
(Dickey \& Lockman, 1990).}
\label{lm-mppow}
\end{center}
\end{figure*}

Data analysis was performed using the software available in the HEAsoft
Package (XIMAGE 3.01, XSELECT v2.0, XSPEC 11.0.1, XRONOS 5.16).
LECS and MECS images displayed a bright pointlike source: 
events for spectral and timing analysis were extracted
from circular regions centered on the source, 
with radii of 8 and 4 arcmin, respectively. 
We extracted LECS and MECS background events in the same way,
from off source regions of the field of view
and checked the constance of the count rates
during the whole observation. In any case, since LECS and MECS
backgrounds are not uniformly distributed across the detectors, 
we choose to use  background files extracted from long blank field exposures,
available from the SDC public ftp site (see Fiore et al., 1999).
The spectral analysis were performed with the XSPEC 11.0.1 package, 
using the updated (01/2000) response matrices.

The PDS was operated in the customary collimator rocking mode, where half
the collimator points at the source and half at the background and they are
switched every 96\,s. The background-subtracted PDS spectrum for
S5\,0716+714 was obtained from the standard pipeline analysis carried
out at the {\it Beppo}SAX Science Data Center.

\subsection{OQ\,530}

Because of the faintness of the source and the short duration of the runs 
the source was not detected by the PDS, as can be seen from Table \ref{tab1}, 
and the $3 \sigma$ upper limit does not add any useful information to the
X-ray spectrum (e.g. for the SED in Fig. 6 below). From the comparison of
LECS and MECS data with  background files we decided to constrain 
further the energy range of our analysis to 0.13--8.7 and 0.13--10 keV
for the 3--4 and 26--27 of March observations, respectively.

For both observations we have poor statistics at low energies
(LECS band) and it is difficult to obtain a reliable fit to the observed 
spectrum. We rebinned the LECS energy channels in order to have more than 25 net
counts in each bin: this is a good compromise between having a sufficient
number of bins and significant number of counts in each of them. 
The MECS  detected a slightly higher number of events, so we rebinned 
them to have 30 net counts in each bin.

We fitted our spectra adopting the galactic interstellar absorption 
$N_{\rm H} = 1.18\times 10^{20}$ cm$^{-2}$, as measured by Dickey \& Lockman (1990).
Fitting the 3--4 March spectrum with a single power law model leaves 
large residuals at low energies 
and the quality of this fit is evidently poor ($\chi^2_r/d.o.f. = 1.64/23$).
Therefore we tried to fit the spectrum with the sum of two power law models
and we obtained a very steep soft X--ray 
spectral index $\alpha_1 = 6.7^{+0.2}_{-1.8}$
and a flatter one towards higher energies $\alpha_2= 0.40\pm{0.2}$
(throughout the paper the errors are at 90\% confidence interval for 
one parameter of interest, i.e. $\Delta \chi^2=2.7$). 
The two power laws cross at $E_b \sim 0.3$ keV. This model gives a better 
fit ($\chi^2_r/d.o.f.=0.96/21$, F--test probability $>99.9\%$), but because
the break is near to the soft limit of our energy range and we have only 
a few counts in this part of the spectrum, we can not really
constrain the soft component of the two power law model.
This soft X--ray excess is probably due to the very steep tail of
the synchrotron component. At energy higher than $\sim 0.5 \, {\rm keV}$, 
instead, we are probably observing an inverse Compton component.

We repeated the procedure for the observation of 26--27 March.
A single power law fits the data better than in the first observation, 
however towards low energies hints of positive residuals are still evident.
We obtain $\alpha = 0.75\pm 0.20$ ($\chi^2_r/d.o.f. = 1.03/21$). 
The measured fluxes between the two observations are quite similar,
$F_{2-10 keV} \sim 10^{-12}$ erg cm$^{-2}$ s$^{-1}$ (see Table \ref{table2}).
We performed the fit also with the sum of two power laws. As in the previous 
observation, in the soft X--ray band we find a very steep component,
with practically the same spectral index $\alpha_1=6.7$, but with even
larger uncertainties (basically we can not constrain this component).
At higher energies the second spectral index is $\alpha_2=0.65 \pm 0.2$.
With this model we obtain a lower $\chi^2_r$ ($\chi^2_r = 0.95$), but an
F--test reveals that the improvement of the fit in this case is only 
marginally significant ($\sim 80\%$). In Table \ref{table2} we report
a summary of our analysis.

Also during the 1999 {\it Beppo}SAX observation, Giommi et al. (2002) have an 
indication of a better fit with a broken power law ($\alpha_1=1.3$,
$\alpha_2=0.4$, and $E_{b}=1.8 $\, {\rm keV}). However, their statistics 
is even lower and the fit improvement is marginally significant. In Table 
\ref{history} we report all the published spectral parameters 
for the OQ\,530 X--ray observations.
The spectral shape of the source seems to be almost constant,
except for the ROSAT observation, that found a steeper spectrum.
This is probably due to a mixture of the soft and hard
components, that the PSPC was not able to disentangle. 
The 1 keV flux density is quite variable; the faintest measured
1 keV flux (by {\it Beppo}SAX, in 1999) is in fact more than
three times weaker than the highest one (ROSAT 1990).

We performed the temporal analysis of our observations in the two bands 
0.2--1.8 keV LECS and 1.8--6 keV MECS. Because of the faintness of the source
we choose a time binning of 1 hour, in order to reduce uncertainties caused
by the poor statistics. During both observations we did not detect variability.

\subsection{S5\,0716+714}

During the {\it Beppo}SAX observation of October 2000, S5\,0716+514 
was detected also by the PDS, so we could perform our analysis on a
wider energy range (0.15--57 keV).

We performed a preliminary analysis on LECS+MECS  and 
MECS+PDS data alone, in order to get some useful hints 
for fitting the whole spectrum.
Initially we fitted the spectra with
single power laws, keeping the absorption parameter 
fixed at the galactic value: $N_{\rm H}=3.9\times10^{20}$ cm$^{-2}$ 
(Dickey \& Lockman, 1990). 
The LECS+MECS spectrum is badly fitted by this model 
($\chi^2_r/d.o.f. = 1.99/76$), with large positive residuals
below 0.5 keV (see Fig. \ref{lm-mppow}). Therefore we fitted it again
with a two power law model. This model fits the data much better 
($\alpha_1=2.5$ $\alpha_2=0.7$, $\chi^2_r/d.o.f.=0.86/74$).
The two power laws cross at $\sim 1$ keV.
The MECS+PDS spectrum is instead fitted quite well by a single power 
law model, with a spectral index somewhat steeper than the second spectral index 
of the LECS+MECS two power law fit ($\alpha=0.9$, $\chi^2_r/d.o.f.= 1.08/36$).
However, we note that due to the weak signal in the PDS, the fit
is clearly dominated by the MECS data. 
A detailed analysis of the PDS light curve in search for spikes gave
negative results. This, together with the lack of other obvious X-ray 
sources inside our field of view, confirm us on the reality of the PDS
detection. We then fitted the whole LECS+MECS+PDS spectrum with a two
power law model, finding a good fit to the data ($\chi^2_r/d.o.f.=0.92/77$).
The two power laws cross at $\sim 1.5$ keV. In Table 2 
we report the best--fit parameters.

\begin{figure}
\centerline{
\vbox{
\psfig{figure=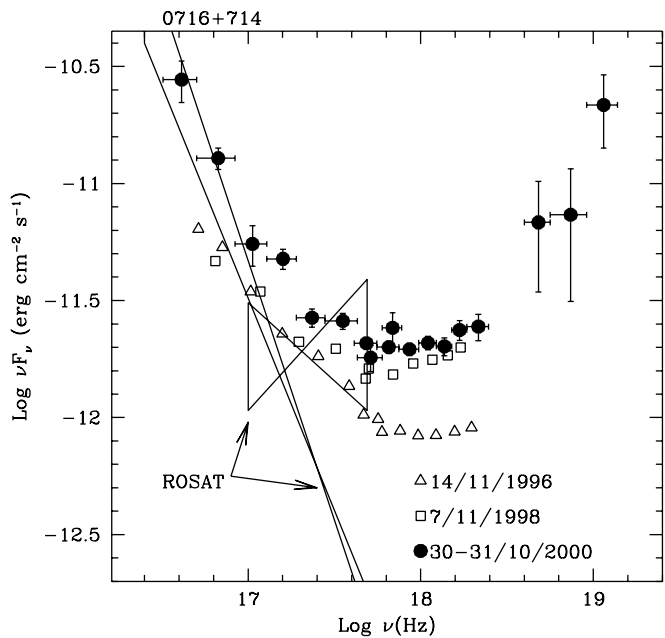,width=10cm}
}}
\vskip -1 true cm
\caption{X--ray spectral energy distribution of S5\,0716+714 during
the October 2000 {\it Beppo}SAX observation (filled circles). 
We plotted also November 1996
(open triangles), November 1998 (open squares) {\it Beppo}SAX
(Giommi et al. 1999) and 1991 ROSAT SEDs (Cappi et al. 1994).}
\label{sedx}
\end{figure}

Although S5 0716+714 was not at its maximum recorded optical luminosity 
($R_C\simeq$12.5), measured a few days before our -X-ray observation,
it was still about 0.8 mag brighter than during the two previous 
{\it Beppo}SAX exposures (Giommi et al. 1999). It was brighter than
the previous observations also in the X--ray band (see Table \ref{history}. 
During our observation the source was in an X--ray state higher
in flux but similar in shape to that detected  by ROSAT in 1991,
when it was revealing again a concave spectrum with an energy break
at $\sim 0.8$ keV (Cappi et al. 1994,  Comastri et al. 1995). 
The historical SED in the X--ray band is displayed in Fig. \ref{sedx}. 
In all observations the transition between the steep soft X--ray component
and the much harder spectrum at higher X--ray energies is clearly detected.
During our observation the soft X--ray spectrum is steeper ($\alpha_1 = 2.4$),
than during the previous {\it Beppo}SAX observations ($\alpha=1.7$ in 1996; 
$\alpha=1.3$ in 1998). This seems to confirm the trend noticed by Giommi et al. 
(1999) of a steeper soft X--ray spectrum when the source is brighter.
\begin{figure}[!b!]
\centerline{
\vbox{
\psfig{figure=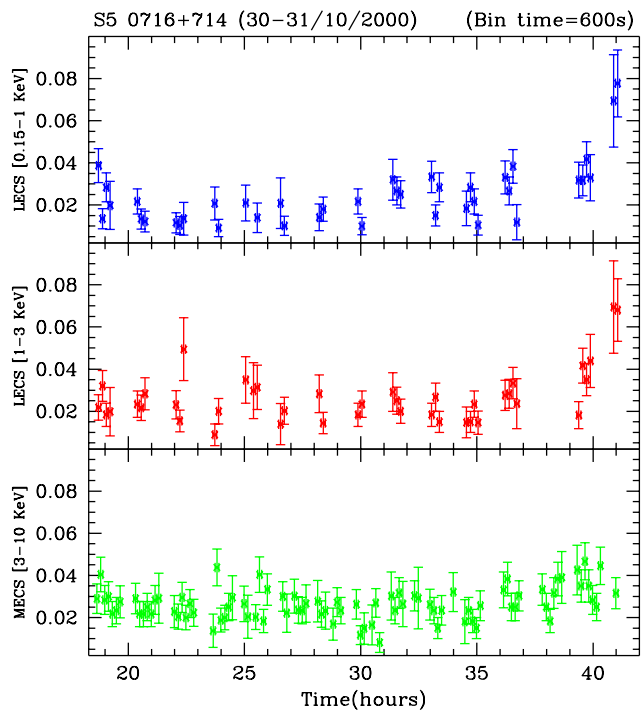,width=10cm}}}
\vskip -0.5 true cm
\caption{LECS 0.15--1 keV (top panel), 1--3 keV (mid panel) and 
MECS 3--10 keV (bottom panel) light curves of S5\,0716+714 during 
the {\it Beppo}SAX observation of October 2000. The time bin is of 10
minutes. Bins with less than $20\%$ effective exposure time have 
been excluded. Time axis is hours starting from October 30, 2002.}
\label{3curve}
\end{figure}

We performed the temporal analysis of our observation in the three 
different energy bands: 0.15--1.0, 1.0--3.0 and 3.0--10 keV.
In Fig. \ref{3curve} we display the two LECS and the higher energy MECS 
light curves. They seem almost constant for the first
20 hours of the observation. Then the two LECS fluxes increase by a 
factor of three on a time scale of one hour. This variation could be
even larger since our measurements stopped before detecting any fading. 
The source displayed a similar behaviour also in the 1.5--3.0 keV
MECS band (not plotted here).
In spite of the formal break that seems to be at around 1.5 keV, the
emission in the two softer energy bands are clearly connected. Clearly,
the soft X--ray component component is still relevant in the 1--3 keV band. 
However, by selecting only the counts above 2 keV we find that the variability
is much less pronounced. The variability is essentially absent in the 3--10
keV band, where the hard X--ray component is dominant (see Fig. \ref{3curve}).
A similar behavior was already detected in a previous {\it Beppo}SAX 
observation (Giommi et al. 1999).

\begin{figure}
\centerline{
\psfig{figure=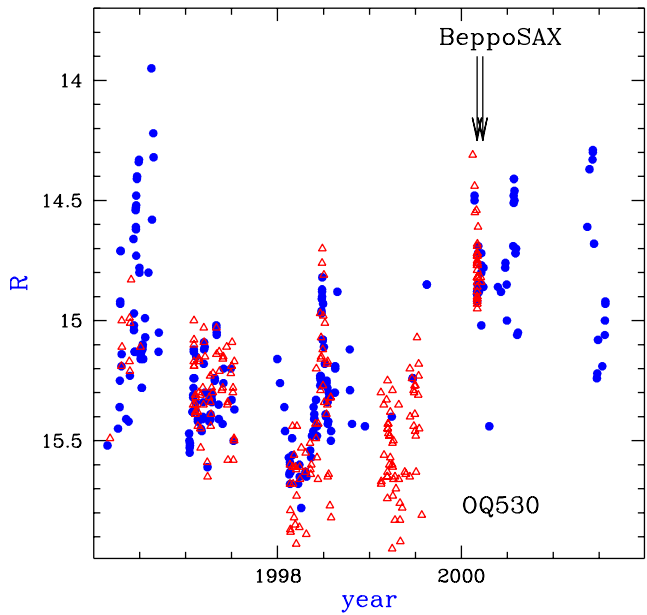,width=10cm}}
\vskip -1 true cm
\caption{The optical, $R_C$ magnitude, light curve of OQ\,530 since February 
1996. Open triangles are data from the Perugia observations (the typical 
error for these measurements is of the order of 0.1 magnitude), filled circles
are from Roma (the typical error for these measurements is of the order of
0.01-0.02 magnitude).
The two arrows indicate the dates of the {\it Beppo}SAX pointings,
activated by a flare--like episode, but carried out when the source 
had already faded to $R_C \sim 15$ mag (see text). }
\label{oq530_lc_ott}
\end{figure}

\section{Optical, Radio and TeV data}

\subsection{OQ\,530}

The optical $R_C$ light curve since February 1996 of the source is shown in 
Fig.\,\ref{oq530_lc_ott}. It is characterized by a large variability with fast
and strong flares
(of the order of half magnitude in a few days). The average trend seems to be
decreasing until the beginning of 1999, when an increasing trend seems to
begin. The average value was $R_C=$15.3
in the period  Jan 1997 -- Aug 1999 (JD 400 -- 1400) so that our observation at 
$R_C=$14.3 on Feb 14, 2000 was clearly indicative of a remarkably
bright state. During the first {\it Beppo}SAX pointing on March 3 the source 
had already dropped at $R_C$=14.9 and remained stable within 0.03 mag; during 
the second one (March 26) the source was about at the same optical luminosity 
but we could observe it only for a short time due to bad weather conditions.
Between the two {\it Beppo}SAX pointings the source remained substantially
constant, with small ($\sim$0.1 mag) variations.
The energy spectral slope ($F_{\nu}\propto \nu^{-\alpha}$) from our 
$BVRI$ observations was 1.63$\pm$0.06, a typical value for the source 
at this luminosity level (D'Amicis et al. 2002). On the other hand, 
the spectral shape (and flux levels)
is affected by the contribution of the host galaxy, and in Fig. 6
the simultaneous optical points have been galaxy--subtracted, using the 
host magnitude given in the Introduction.

Very Long Baseline Interferometric radio observations with EVN at 5 GHz were
performed one year before (February 1999) and one year after (June 5, 2001)
the {\it Beppo}SAX pointing by some of us (Massaro et al. 2002). The source
showed a core--jet structure, already detected in 1996 (Fomalont et al. 2000):
the core had a flux density of 440 mJy at that epoch, but dropped to 204 mJy
on February 1999 and rose again up to 685 mJy on June 2001. This increase,
larger than a factor of 3, of the core radio brightness supports the 
development of strong activity following the optical burst of February 2000.

OQ\,530 has been pointed in the TeV band by the HEGRA telescopes from
July 1 to July 16 1999, and on March 5 2000.
The total exposure time is of 9.7 hours, a quarter of which is from
the March observation (1 day after the {\it Beppo}SAX observation).
The source was never detected with a total upper limit  UL(99\%, 
E$\, > 1.1$ TeV) $= 1.35 \times 10^{-12} \ {\rm photons \ cm^{-2} \ s^{-1}}$ 
(Aharonian et al., HEGRA collaboration, 2002, in preparation).

\subsection{S5\,0716+714}

As stated in the Introduction, the source was very bright the week before 
the {\it Beppo}SAX observation: $R_C = 12.55 \pm 0.02$ mag on
October 22 and 12.58$\pm$0.03 on October 28. Only few optical photometric 
data, taken during the X--ray pointing, are available, but these are enough 
to derive the SED and to compare it with that of previous observations. 
In particular, $R_C$ and $I_C$ were measured equal to 
13.05$\pm$0.04 and 12.55$\pm$0.04 mag at 22:40 UT (October 30) with the 0.5m
reflector of the Astronomical Station of Vallinfreda (Rome). 
On November 1 22:10 UT its luminosity has faded at $R_C$=13.31 
(Vallinfreda), indicating that over the entire period S5\,0716+714 showed
brightness variations on intraday time scale with amplitude of a few 
tenths of magnitude, as is usually observed in this source (Nesci et al. 2002).

\begin{figure}
\centerline{
\psfig{figure=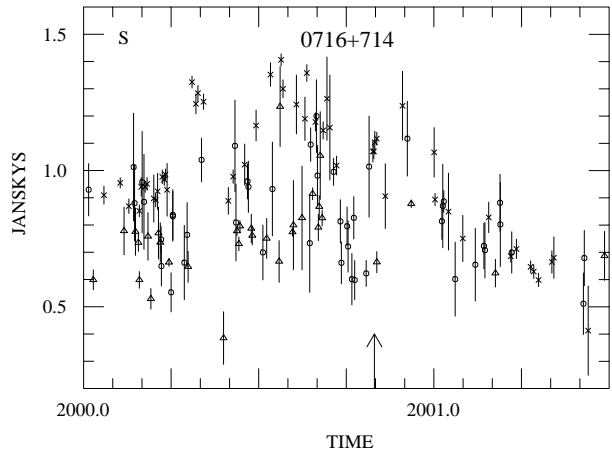,angle=0,width=9.5cm}}
\caption{The 1.5 years light curve of S5\,0716+714 at three radio frequencies:
triangles are data at 4.8 GHz, open circles data at 8.0 GHz, and x denotes
data at 14.5 GHz. The measurements are given as daily averages. The source
show erratic variability during all year 2000, then there seem to be
a steady decline in the fluxes.
}
\label{0716_lc_rad}
\end{figure}

During the {\it Beppo}SAX observation the radio flux of S5\,0716+714 was 
monitored at the UMRAO, using the procedures 
and calibrators described in Aller et al. (1985). 
At 14.5 GHz, from October 28 to November 1, the flux density 
remained constant with a mean value of 1.1 Jy. Other measurements at
8.0 GHz, performed from October 1 to 21 give a flux density in the
range $0.60 - 0.85 \pm 0.1$ Jy. 
In Fig.\,\ref{0716_lc_rad} we plot the 1.5 years radio light curve of S5\,0716+714
at three frequencies. The source seems to be in an active state up to the end of
year 2000, then there seems to be a steady decline in the fluxes.

During our {\it Beppo}SAX observation S5\,0716+714 has been observed
also in the TeV band by the HEGRA telescopes for about 2 hours on October 
31 2000. However, the source has not been detected with an upper limit 
UL(99\%, E$\, > 1.7$ TeV) $= 2.5 \times 10^{-12} \ {\rm photons \ cm^{-2} \ s^{-1}}$ 
(Aharonian et al., HEGRA collaboration, 2002, in preparation).

\begin{table*}[!t!]
\begin{center}
\begin{tabular}{ccccccc}
\multicolumn{7}{c}{\bf{OQ\,530}}\\
\hline
\hline
Date & Mission & $\alpha_1$ & $\alpha_2$ & F$_{1 keV}$ & F$_{2-10 keV}$ & Ref.\\
 & & &  & $\mu$Jy & $\times10^{-12}$ erg cm$^{-2}$ s$^{-1}$ & \\
\hline
\hline
December 1980 & Einstein & $0.48^{+1.0}_{-0.3}$ & & 0.21 
& & 1\\
\hline
July 1990 & ROSAT & $1.04\pm0.05$ & & 0.32 & & 2\\
\hline
February 1999 & {\it Beppo}SAX &$0.55^{+0.27}_{-0.32}$ & & 0.09 
& 0.67 & 3\\
\hline
3-4 March 2000 & {\it Beppo}SAX & $0.55\pm0.2$ & & 0.14 & 1.1 & \\
\hline
26-27 March 2000 &  {\it Beppo}SAX & $0.75\pm0.2$ & & 0.16 & 0.9 & \\
\hline

\multicolumn{7}{c} {}\\
\multicolumn{7}{c}{\bf{S5\,0716+714}}\\
\hline
\hline
8 March 1991 & ROSAT & $2.67\pm0.12$ & $1.00$ & 0.78 & & 4\\
\hline  
14 November 1996 & {\it Beppo}SAX & $1.7\pm0.3$ & $0.96\pm0.15$ & 0.13
      & $1.4$ & 5\\
\hline
 7 November 1998 & {\it Beppo}SAX & $1.3\pm0.4$ & $0.73\pm0.18$ & 0.19
      & 2.6 & 5\\
\hline
30-31 October 2000 & {\it Beppo}SAX & $2.40^{+0.4}_{-0.3}$ & $0.60^{+0.25}_{-0.35}$ 
      & 1.1  & 3.3 & \\
\hline
\end{tabular}
\caption{OQ\,530 and S5\,0716+714 X--ray spectral parameters.
(1) Worrall \& Wilkes (1990); (2) Comastri et al., 1995; (3) Padovani et al., 
in prep.; (4) Cappi et al. (1994); (5) Giommi et al. (1999). }
\label{history}
\end{center}
\end{table*}

\section{The Spectral Energy Distributions}

We have used a homogeneous, one--zone synchrotron inverse Compton model 
to reproduce the SEDs of our sources. The model is very similar to the
one described in detail in Spada et al. (2001), it is the ``one--zone''
version of it. Further details can be found in Ghisellini, Celotti \&
Costamante (2002), where the same model has been applied to less powerful
BL Lacs. The main assumptions of the model are: 

\begin{itemize}

\item The source is cylindrical, of radius $R$ and
thickness $\Delta R^\prime = R/\Gamma$ (in the comoving frame, where 
$\Gamma$ is the bulk Lorentz factor).

\item The source is assumed to emit an intrinsic luminosity $L^\prime$ 
and to be observed at a viewing angle $\theta$ with respect to the jet axis. 

\item The particle distribution $N(\gamma)$
is assumed to have the slope $n$ [$N(\gamma)\propto \gamma^{-n}$] above the
random Lorentz factor $\gamma_{\rm c}$, for which the radiative (synchrotron 
and inverse Compton) 
cooling time equals $\Delta R^\prime /c$.
The motivation behind this choice is the assumption that 
relativistic particles are injected in the emitting volume 
for a finite time, which we take roughly equal to the light 
crossing time of the shell. This crossing time is roughly equal to the
time needed for two shells to cross, if they have  
bulk Lorentz factor differing by a factor around two.
The electron distribution
is assumed to cut--off abruptly at $\gamma_{\rm max}>\gamma_{\rm c}$.
We then assume that between some $\gamma_{\rm min}$ and $\gamma_{\rm c}$ 
the particle distribution $N(\gamma)\propto \gamma^{-(n-1)}$.
This choice corresponds to the case in which the injected particle distribution
is a power law ($\propto \gamma^{-(n-1)}$)
between $\gamma_{\rm min}$ and $\gamma_{\rm max}$, with 
$\gamma_{\rm min}<\gamma_{\rm c}$.
Below $\gamma_{\rm min}$ we assume $N(\gamma)\propto \gamma^{-1}$.
The value of $\gamma_{\rm max}$ is not crucial, due to the fact
that the electron distribution, for our sources, is steep.
It has been chosen to be a factor $\sim 100$ larger than $\gamma_{\rm min}$.

\end{itemize}

The simultaneous optical observations help in defining the 
low frequency peak of the SED, which in both sources
must be located at frequencies lower than the optical. 
This helps to constrain the values of the input parameters of our model.
Since we determine $\gamma_{\rm peak}$ as the energy of those electrons
that can cool in the injection time (i.e. $\gamma_{\rm peak}=\gamma_{\rm c}$), 
there is a relation between $\nu_{\rm peak}$ and the cooling rate.
If the latter is dominated by synchrotron emission, the location
of $\nu_{\rm peak}$ constrains the value of the magnetic field.
This in turn fixes the amount of inverse Compton radiation.
According to our simultaneous data, the slope of the optical emission
for both sources is steep (i.e. $\alpha>1$), requiring 
the synchrotron part of the spectrum to peak below the optical band.
On the other hand, the infrared data, even if not simultaneous,
suggest that $\nu_{\rm peak}$ cannot be far below the optical.
This motivates the choice of the adopted $\nu_{\rm peak}$
as listed in Table 4.

The remaining degree of freedom for the choice of the input parameters is 
due to the value of the beaming factor (i.e. $\Gamma$ and the viewing 
angle $\theta$), and to the redshift.
Another important input parameter is the size of the emitting region.
The extremely rapid variability shown by S5 0716+715 during our observations 
refers to the soft X--ray flux, while timescales of the order of a $\sim$day 
are typical in the optical (see e.g. Ghisellini et al. 1999), at frequencies closer 
to the synchrotron peak.
The size of the emitting region is therefore constrained to be less than one light--hour
or a light--day (i.e. $R\le ct_{\rm var} \delta/(1+z)$).
Although a one--zone homogeneous model is forced to use the minimum variability
timescale observed at any band to constrain the size, it is also clear that
this model is a simplification of a scenario which may be more complex.
Our choice of $R=2\times 10^{16}$ and $\delta\sim 17$ corresponds to a minimum variability
timescale of $t_{var}\sim 11(1+z)$ hours.
Assuming a more compact homogeneous source, while more in agreement with the
soft--ray variability, implies a more dominant self Compton component, 
unless a larger beaming factor is also assumed.
For the redshift, since only the lower limit $z=0.3$ is known for S5\,0716+714,
we have assumed $z=0.3$ in all our calculations for S5 0716+714 but one case,
for which $z=1$ has been assumed, in order to check what are the output parameters
mostly affected by the choice of a particular redshift.

We have tried to reproduce the three SEDs of S5 0716+714 by changing the
minimum possible number of input parameters: the most significant change
is in the injected luminosity, which in Oct 2000 is found to be twice as much 
as in 1996 and 1998.

An important consequence for our modeling is that the SSC component alone 
cannot reproduce the observed emission in the EGRET band in S5 0716+714,
as shown in the bottom panel of Fig. 7. Since these $\gamma$--ray data are
not simultaneous, there is in principle no need to reproduce them. 
However, since the source should be in a bright state, it may be likely that
the level of $\gamma$--ray emission during our campaign was similar to what 
EGRET previously found (Lin et al. 1995). If we want to account for a flux
in the 0.1--10 GeV band similar to what found by EGRET, we should then consider
the possibility that the high energy emission can be 
produced by the scattering of external soft photons, either produced in a putative 
broad line region or in the accretion flow.                          
We have allowed for some external Compton radiation 
by assuming the presence of external photons produced by, e.g.
a broad line region of relatively low luminosity, as listed in Table 4.
The effect of this mechanism is to produce the second ``hump" at the largest
$\gamma$--ray energies, as shown by the solid line in Fig. 6 (for OQ 530) and
by the top panel of Fig. \ref{0716_sed} (for S5 0716+714). This high energy
bump is due to the fact that the seed photons due to the emission lines, 
in the comoving frame, are at the typical frequency of $10^{15}\Gamma$ Hz, 
larger than the synchrotron peak frequency in the same frame, 
which is $\nu_{\rm peak}/\delta \sim \nu_{\rm peak}/\Gamma$).

The possibility of a broad line region in objects such as S5\,0716+714 is 
questionable: the line emission could be variable (as in BL Lac itself),
or always very weak (as seems to be the case for low power, high frequency
peaked BL Lacs).
On the other hand, to produce extra inverse Compton emission,
other processes could be important, as in the ``mirror model"
discussed by Ghisellini \& Madau (1996). Therefore the values listed
in Table \ref{tab_sed}, which formally refer to the broad line region,
could instead be appropriate for other possible contributors of seed
photons. Note that for OQ\,530 there are only upper limits for the
$\gamma$--ray flux, which are not stringent enough to constrain the
presence of any external component (compare the solid and dashed lines).

\begin{figure}
\psfig{figure=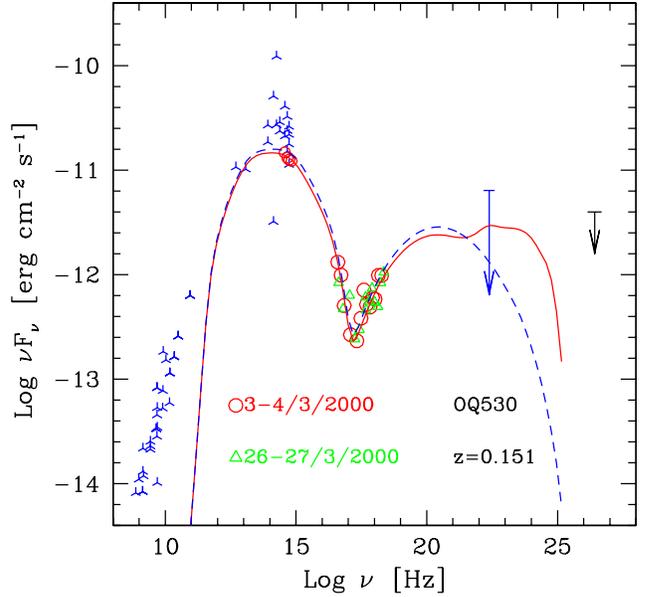,width=9.5cm}
\vskip -0.5 true cm
\caption{The SED of OQ\,530. Star symbols are from the NASA Extragalagtic 
Database (NED). The solid line refers to the SSC plus EC model, while
the dashed line corresponds to a pure SSC models, with no contributions
from seed photons generated externally to the jet. All other parameters
of the two models are the same.}
\label{oq530_sed}
\end{figure} 
\begin{figure}
\psfig{figure=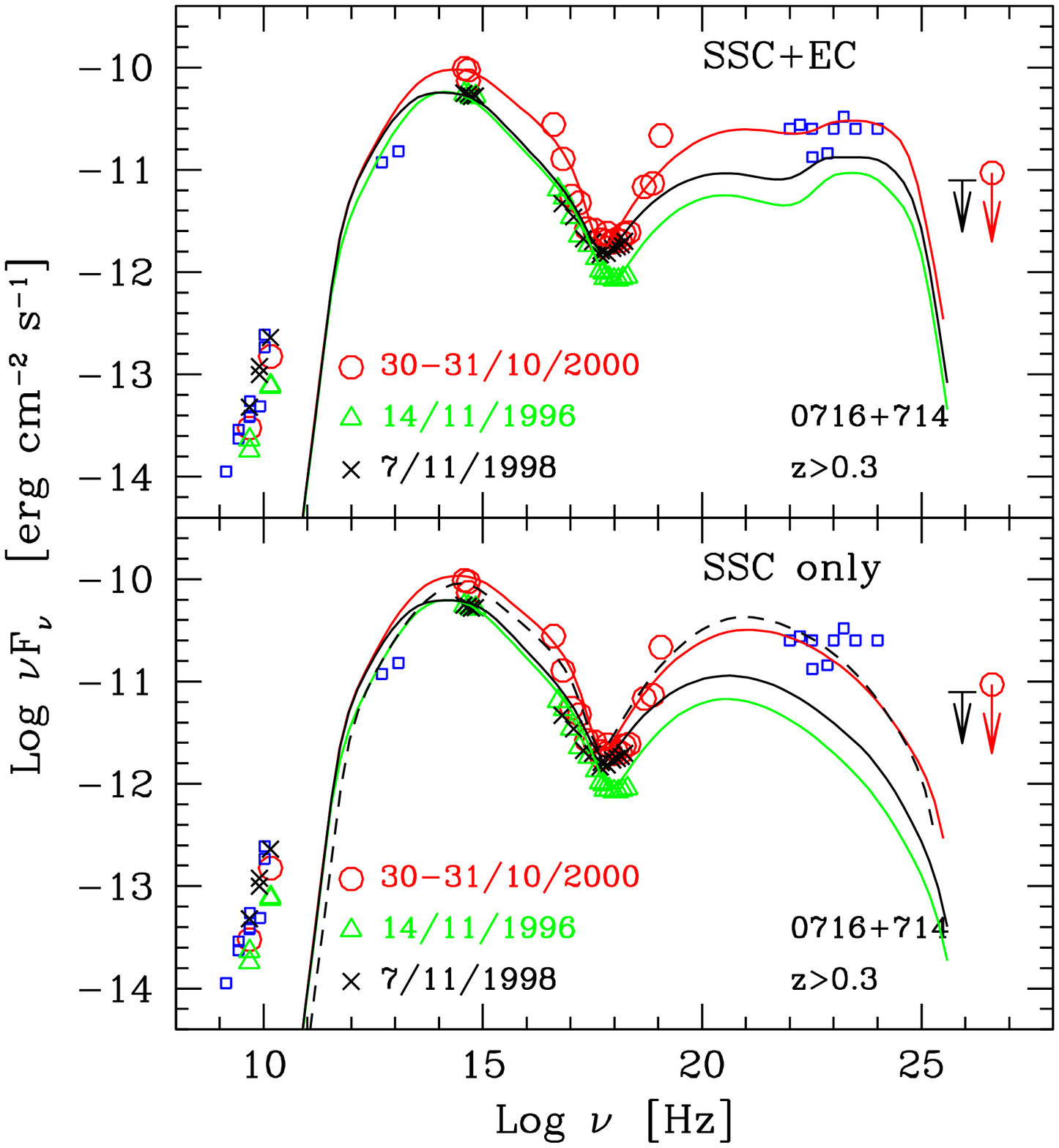,width=9.5cm,height=12 true cm}
\caption{The SED of S5\,0716+714. In the top panel we show as solid
lines the SSC plus external Compton model we have applied to the three
labeled observing campaigns. For all models $z=0.3$ was assumed.
The bottom panel shows pure SSC models, with no contributions from seed
photons generated externally to the jet. The other input parameters are
the same (solid lines), while $z=1$ is assumed for the dashed line.
The square symbols refer to not simultaneous observations taken from
NED and from Lin et al. (1995).}
\label{0716_sed}
\end{figure}

The input parameters are listed in Table \ref{tab_sed} and the models are
shown in Fig. \ref{oq530_sed} and Fig. \ref{0716_sed}.
The applied model is aimed at reproducing the spectrum originating 
in a limited part of the jet, thought to be responsible for most of 
the emission. 
This region is necessarily compact, since it must account for the 
fast variability shown by all blazars, especially at high frequencies. 
The radio emission from this compact region is strongly self--absorbed,
and thus the model cannot account for the observed radio flux. 
This explains why the radio data are systematically above the model 
fits in the figures.

\begin{table*}
\begin{center}
\caption{{
Model Input Parameters. 
Column (1): name of the source;
column (2): observation date;
column (3): intrinsic luminosity $L^\prime$; 
column (4): luminosity of broad emission lines $L_{\rm BLR}$: the first value
(i.e. zero) corresponds to assume no contribution of external photons to the formation 
of the IC spectrum; 
column (5): dimension of the broad line region $R_{\rm BLR}$;
column (6): magnetic field $B$;
column (7): size of the emitting region $R$;
column (8): bulk Lorentz factor $\Gamma$; 
column (9): viewing angle $\theta$ (in degrees);
column (10): beaming factor $\delta$; 
column (11): slope of the particle distribution $n$;
column (12): minimum Lorentz factor of the injected electrons $\gamma_{\rm min}$;
column (13): Lorentz factor of the electron emitting at the peaks, $\gamma_{\rm peak}$;
column (14): synchrotron peak frequency $\nu_{\rm peak}$.}
Note that $\gamma_{\rm peak}$ and $\nu_{\rm peak}$ are derived quantities and 
not input parameters.
}
\label{tab_sed}
\begin{tabular}{llllllllllllll}
\hline
Name &date  &$L^\prime$    &$L_{\rm BLR}$ &$R_{\rm BLR}$ &$B$ &$R$  &$\Gamma$ &$\theta$ 
   &$\delta$ &$n$  
&$\gamma_{\rm min}$  &$\gamma_{\rm peak}$ &$\nu_{\rm peak}$ \\
     &       & erg s$^{-1}$ &erg s$^{-1}$ & cm  &G    & cm    &     &  &   &     
    &      &      &Hz \\
\hline
0716 ($z$=0.3) &Oct 2000 &5.0e42  &0--5e42  &3e17  &2.5  &2e16 &15  &3.4 &16.7 &3.7  
    &5.0e2 &1.8e3 &5.1e14 \\ 
0716 ($z$=0.3) &Nov 1998 &2.7e42  &0--5e42  &3e17  &2.5  &2e16 &15  &3.4 &16.7 &3.9  
    &4.0e2 &2.0e3 &6.2e14 \\ 
0716 ($z$=0.3) &Nov 1996 &2.2e42  &0--5e42  &3e17  &3.0  &2e16 &15  &3.4 &16.7 &3.9  
    &5.0e2 &1.5e3 &4.2e14 \\ 
0716 ($z$=1)   &Oct 2000 &4.0e43  &---      &---   &5    &2e16 &15  &3   &18.5 &3.7  
    &1.2e3 &1.2e3 &5.0e14 \\ 
OQ\,530        &Mar 2000 &8.0e41  &0--5e42  &2e17  &2.8  &1e16 &10  &5.0  &11.4 &3.9  
    &3.3e2 &2.2e3 &5.8e14 \\ 
\hline  
\end{tabular}
\end{center}
\label{tab_sed}
\end{table*}

\begin{figure}
\psfig{figure=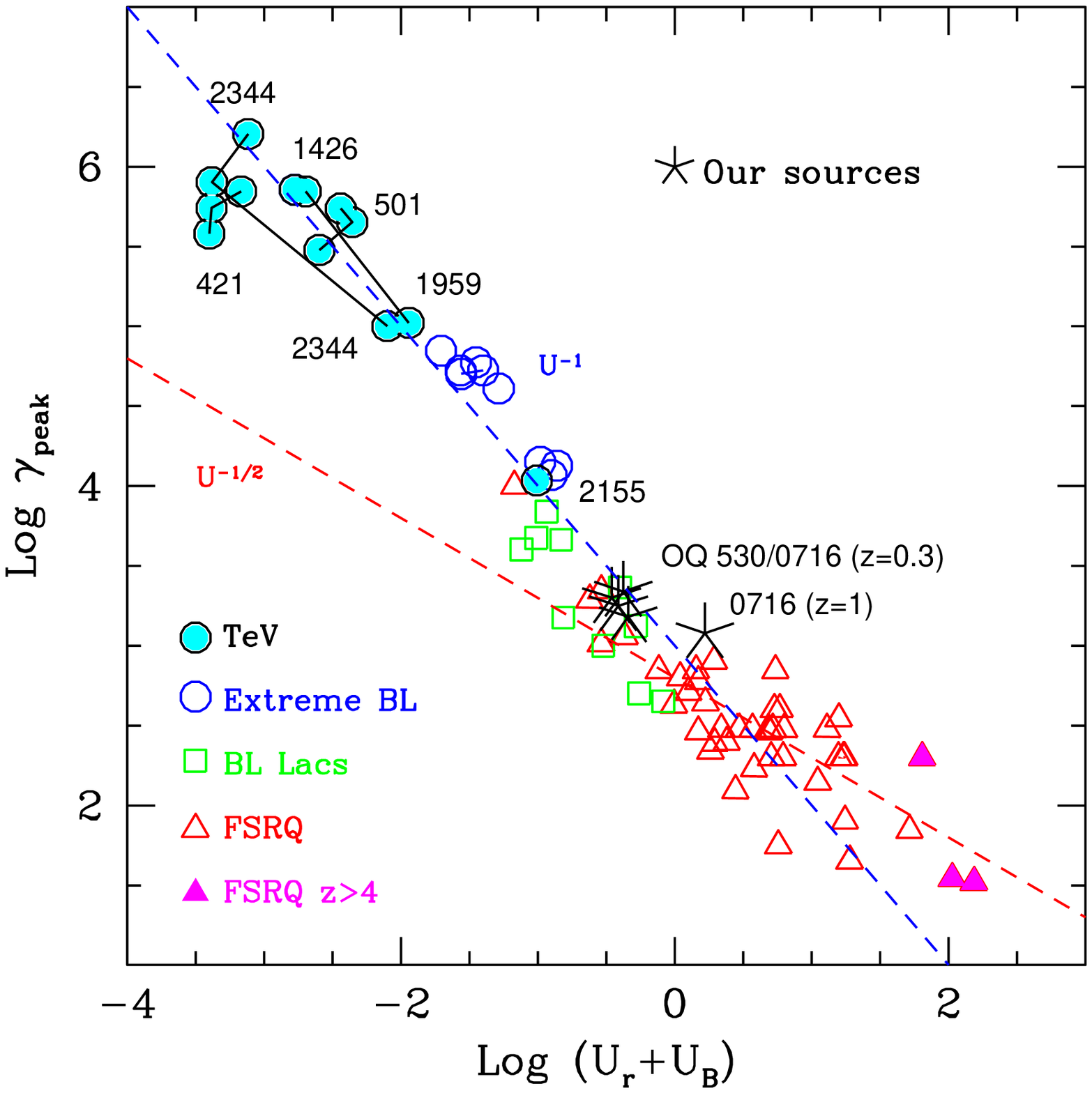,width=9.5cm}
\vskip -0.5 true cm
\caption{Location of our sources in the $\gamma_{\rm peak}$--$U$
plane, where $U$ is the sum of the radiation and magnetic energy
densities, as measured in the comoving frame. Note that our two
sources are located in the region where the two branches 
(proposed by Ghisellini, Celotti \& Costamante, 2002) join,
supporting the intermediate nature of these two BL Lacs
(for S5\,0716+714 we report also one case with $z=1$). 
Adapted from Ghisellini, Celotti \& Costamante (2002).
}
\label{corr}
\end{figure} 

\section{Discussion}

We presented the results of {\it Beppo}SAX ToO observations
of two BL Lac objects observed while they were in a high state.
Once again these observations were triggered from optical 
monitoring. In our ToO program we have probably been biased
toward sources that have an higher optical variability.
These should be the blazars that have the synchrotron peak 
in the IR-optical band and these are of course essentially LBL
or intermediate blazars. In particular the two sources studied 
in this paper belong to the intermediate BL Lac class in terms
of their spectral energy distribution.
In this regard they resembles ON 231 (Tagliaferri et al. 2000):
all these three BL Lacs show evidence for a concave 
X--ray spectrum in the 0.1--10 keV band, a signature
of the presence of both the steep tail of the synchrotron
emission and the flat part of the inverse Compton spectrum.
Another source that we studied more than once and in which
we detected both components, with a concave X--ray spectrum,
or only the Compton component, is BL Lac itself (Ravasio et al. 2002).
A common result that we found in ON\,231, BL\,Lac and S5\,0716+714
is that fast variability was detected for all the three sources only 
in the synchrotron component. 
This can be interpreted with the
presence in the X--ray band of a Compton component (slowly variable on
time scale of months), and the tail of a synchrotron component with fast and
the erratic variability.

The Compton emission we see in the X--ray band is well below the Compton peak 
and it is produced by low energy electrons scattering low frequency synchrotron
photons. 
The short timescale variability seen in the synchrotron part can be reproduced
by changing the slope of the injected electron distribution, without
affecting the total injected power. This is in line also with the fact
that we do not see large shifts of the synchrotron peak frequency during 
our observations, that are usually performed when the source is in a higher 
state than the previous observations (e.g. Fig. \ref{0716_sed}, see Tab. 4).

In the framework of the blazar sequence proposed by
Fossati et al. (1998) the two sources studied here have bolometric
luminosities intermediate between HBLs (with bolometric luminosities
$L \sim 10^{45}$ erg s$^{-1}$) and powerful broad line blazars 
(with bolometric luminosities up to 
$L\sim 10^{48}-10^{49}$ erg s$^{-1}$,
see Fig. 12 in Fossati et al. 1998). 
In the case of S5\,0716+714, assuming $z=0.3$  and 
a cosmology with $H_0=65$ km s$^{-1}$ Mpc$^{-1}$, $\Omega_m=0.3$ and
$\Omega_{\Lambda}=0.7$, the bolometric luminosity is 
$\sim (3-5)\times 10^{46}$ erg s$^{-1}$,
while in the case of OQ 530 we have 
$(2-10)\times 10^{45}$ erg s$^{-1}$.
The values of their $\gamma_{\rm peak}$ 
and the comoving radiation plus magnetic energy density
locate our objects along the correlation found
by Ghisellini et al. (1998) and Ghisellini, Celotti \& Costamante 
(2002), as shown in Fig. \ref{corr}.
Notice also that the two objects are located at the joining point
between the two branches (i.e $\gamma_{\rm peak} \propto U^{-1}$
and  $\gamma_{\rm peak} \propto U^{-1/2}$) proposed by Ghisellini,
Celotti \& Costamante (2002) as characterizing all blazars
(we also report for comparison one case with $z=1$
for S5\,0716+714). 
Note also that the 3 states of S5\,0716+714 studied here
correspond to this object ``moving" along the 
$\gamma_{\rm peak}$--$U$ correlation, even if the small dynamic
range of the variations makes this result tentative at best.

Our results show the importance of monitoring these 
variable sources and of performing multiwavelength observations,
including a X--ray energy band as wide as possible, while 
they are either in a high or in weak state, to obtain the largest
information on the dynamical evolution of the emission components.

\begin{acknowledgements}

We thank the HEGRA collaboration group for letting us have their
results in advance of publication and for useful discussion.
This research was financially supported by the Italian
Space Agency and by the Italian Ministry for University and Research under the
grant Cofin 2001/028773. 
We thank the {\it Beppo}SAX Science Data Center (SDC) for their 
support in the data analysis.
The University of Michigan Radio Astronomy Observatory was funded in part
by the NSF and by the University of Michigan Department of Astronomy.
This research made use of the NASA/IPAC Extragalactic Database (NED)
which is operated by the Jet Propulsion Laboratory, Caltech, under
contract with the National Aeronautics and Space Administration.

\end{acknowledgements}

\end{document}